\documentclass[prl,twocolumn,amssymb,showpacs]{revtex4}

\usepackage{graphicx}
\usepackage{color}
\usepackage{epsfig}
\usepackage{latexsym}
\usepackage{bm}
\usepackage{ulem}

\begin{document}
\renewcommand{\ni}{{\noindent}}
\newcommand{\dprime}{{\prime\prime}}
\newcommand{\be}{\begin{equation}}
\newcommand{\ee}{\end{equation}}
\newcommand{\bea}{\begin{eqnarray}}
\newcommand{\eea}{\end{eqnarray}}
\newcommand{\nn}{\nonumber}
\newcommand{\bk}{{\bf k}}
\newcommand{\bQ}{{\bf Q}}
\newcommand{\q}{{\bf q}}
\newcommand{\s}{{\bf s}}
\newcommand{\bN}{{\bf \nabla}}
\newcommand{\bA}{{\bf A}}
\newcommand{\bE}{{\bf E}}
\newcommand{\bj}{{\bf j}}
\newcommand{\bJ}{{\bf J}}
\newcommand{\bs}{{\bf v}_s}
\newcommand{\bn}{{\bf v}_n}
\newcommand{\bv}{{\bf v}}
\newcommand{\la}{\langle}
\newcommand{\ra}{\rangle}
\newcommand{\dg}{\dagger}
\newcommand{\br}{{\bf{r}}}
\newcommand{\brp}{{\bf{r}^\prime}}
\newcommand{\bq}{{\bf{q}}}
\newcommand{\hx}{\hat{\bf x}}
\newcommand{\hy}{\hat{\bf y}}
\newcommand{\bS}{{\bf S}}
\newcommand{\cU}{{\cal U}}
\newcommand{\cD}{{\cal D}}
\newcommand{\bR}{{\bf R}}
\newcommand{\pll}{\parallel}
\newcommand{\sumr}{\sum_{\vr}}
\newcommand{\cP}{{\cal P}}
\newcommand{\cQ}{{\cal Q}}
\newcommand{\cS}{{\cal S}}
\newcommand{\ua}{\uparrow}
\newcommand{\da}{\downarrow}
\newcommand{\red}{\textcolor {red}}
\newcommand{\blu}{\textcolor {blue}}

\def\lsim {\protect \raisebox{-0.75ex}[-1.5ex]{$\;\stackrel{<}{\sim}\;$}}
\def\gsim {\protect \raisebox{-0.75ex}[-1.5ex]{$\;\stackrel{>}{\sim}\;$}}
\def\lsimeq {\protect \raisebox{-0.75ex}[-1.5ex]{$\;\stackrel{<}{\simeq}\;$}}
\def\gsimeq {\protect \raisebox{-0.75ex}[-1.5ex]{$\;\stackrel{>}{\simeq}\;$}}


\title{  Gammalike mass distributions and mass fluctuations in conserved-mass transport processes }

\author{Sayani Chatterjee$^{1}$, Punyabrata Pradhan$^{1}$ and P. K.
Mohanty$^{2}$}

\affiliation{ $^1$Department of Theoretical Sciences, S. N. Bose National
Centre for Basic Sciences, Block - JD, Sector - III, Salt Lake, Kolkata 
700098, India \\ $^2$CMP Division, Saha
Institute of Nuclear Physics, 1/AF Bidhan Nagar, Kolkata 700064, India}

\begin{abstract}
\noindent{
We show that, in conserved-mass transport processes, the steady-state
distribution of mass in a subsystem is uniquely determined  from the
functional dependence  of variance of the subsystem mass on its mean,
provided that joint mass distribution of subsystems is factorized
in the thermodynamic limit. The factorization condition is not too
restrictive as it would hold in systems with short-ranged spatial
correlations. To demonstrate the result, we revisit a broad class of
mass transport models and  its generic variants, and show that the
variance of subsystem mass in these models is proportional to square
of its mean.
This particular functional form of the variance constrains the
subsystem mass distribution to be a gamma distribution irrespective of
the dynamical rules.
}
\typeout{polish abstract}
\end{abstract}


\maketitle

{\it Introduction.} -- Understanding fluctuations is  fundamental to the
formulation of statistical mechanics. Unlike in equilibrium, where
fluctuations are obtained from  the Boltzmann distribution, there is no
unified principle  to  characterize fluctuations in nonequilibrium.
In this Letter, we provide a   statistical mechanics framework
to characterize steady-state mass fluctuations in
conserved-mass transport processes.

Nonequilibrium processes of mass transport which
happen through fragmentation, diffusion and coalescence are ubiquitous
in nature, e.g., in clouds \cite{Friedlander}, fluids condensing on
cold surfaces \cite{Meakin}, suspensions of colloid-particles \cite{White},
polymer gels \cite{Ziff}, etc. To study these processes, various models
with discrete as well as continuous time  dynamics have been proposed on a
lattice where total mass is conserved \cite{TakayasuPRL1989, MajumdarPRL1998,
KrugGarcia2000, RajeshMajumdar2000, Zia_etal_JPhysA2004, Evans_JPhysA2005,
Evans_PRL2006, Zielen_JSP2002,
Zielen_JSP2003, Mohanty_JSTAT2012}. These models, a paradigm in
nonequilibrium statistical mechanics, are relevant not only for transport of
mass, but can also describe seemingly different nonequilibrium phenomena, as
diverse as dynamics of driven interacting particles on a ring
\cite{KrugGarcia2000}, force fluctuations in granular beads
\cite{Majumdar_Science1995, Majumdar_PRE1996}, distribution of wealth
\cite{Yakovenko, Patriarca_EPJB2010}, energy transport in solids
\cite{KMP1982}, traffic flow \cite{Krauss1996, Chowdhury2000}, and river
network \cite{Scheidegger}, etc.

A striking common feature   in many of these processes is that the
probability distributions of mass at a single site are  described
by {\it gamma distributions} \cite{Majumdar_PRE1996,
KrugGarcia2000, RajeshMajumdar2000, Zielen_JSP2002,
Zielen_JSP2003, Mohanty_JSTAT2012}. In several other cases, e.g.,
in cases of wealth distribution in a population \cite{Yakovenko,
Patriarca_EPJB2010, AnirbanC} or force distribution in granular
beads \cite{Majumdar_Science1995, Majumdar_PRE1996}, the
distribution functions are not always exactly known, but
remarkably they can often be well approximated by gamma
distributions. Although these models have been studied intensively 
in the past decades, an intriguing question
\cite{Bertin_PRL2004} - why the gamma-like distributions arise in
different contexts irrespective of different dynamical rules -
still remains unanswered.

In this Letter, we address these issues in general and explain in
particular why  mass-transport processes often exhibit gamma-like 
distributions. 
Our main result is that, in the thermodynamic limit, the functional
dependence of variance of subsystem mass on its mean 
uniquely determines the probability distribution of the subsystem mass, 
provided that (i) total mass is conserved and (ii) the joint 
probability distribution of masses in subsystems has a factorized form 
as given in Eq. \ref{PC1}. In other words, if the conditions (i)
and (ii) are satisfied, the probability distribution   $P_v(m)$ 
of mass $m$ in a subsystem of size $v$
can be determined from the functional form of the variance $\sigma_v^2
\equiv \psi(\langle m \rangle)$ where $\langle m \rangle$ the mean.  
In fact, $\psi(\langle m\rangle)$ in systems with short-ranged spatial 
correlations can be calculated by integrating two-point spatial 
correlation function. An important  consequence of the main result is 
the following. When the variance of subsystem mass  is
proportional to the square of its mean, i.e., 
$\psi(\langle m \rangle)= {\langle m \rangle^2}/{v \eta}$ 
with a parameter $\eta$ that
depends on the  dynamical rules of a particular model, the subsystem mass
distribution is a {\it gamma} distribution, \be P_v(m) =
\frac{1}{\Gamma(v\eta)} \left( \frac{v\eta}{\langle m \rangle}
\right)^{v\eta} m^{v\eta - 1} e^{-v \eta m/\langle m \rangle},
\label{Pm1} \ee where $\Gamma(\eta)=\int_0^{\infty} m^{\eta-1}
\exp(-m) dm$ the gamma function.
Indeed, we find that $\psi(\langle m \rangle)$ is proportional   to 
${\langle m \rangle^2}$ in  a broad class of mass-transport models, 
which explains why  these  models  exhibit gamma distributions.

It might be  surprising how the variance alone could determine
the probability distribution $P_v(m)$ as an analytic probability
distribution function is  uniquely determined only if {\it  all} 
its moments are provided. However, the result can be understood
from the fact that, for a system satisfying the above conditions
(i) and (ii), there exists an equilibrium-like chemical potential
and consequently a fluctuation-response relation that relates mass
fluctuation  to the response due to a change in chemical potential.
This relation, analogous to equilibrium fluctuation-dissipation theorem,
provides a unique functional dependence of the chemical potential on mean
mass and constrains  $P_v(m)$ to take a  specific form.

{\it Proof.} -- Let us consider a mass-transport process on a
lattice of $V$ sites with continuous mass variables $m_i \ge 0$ at
site $i=1, \dots, V$. With some specified rates, masses get
fragmented and then the neighboring fragments of mass coalesce
with each other. At this stage, we need not specify details of the
dynamical rules, only assume that the total mass $M=\sum_{i=1}^V
m_i$ is conserved. We partition the system into $\nu$ subsystems
of  equal sizes $v=V/\nu$ and consider fluctuation of mass $M_k$
in $k$th subsystem. We assume that the joint probability ${\cal
P}(\{M_k\})$ of subsystems having masses $\{M_1, M_2, \dots
M_{\nu} \} \equiv \{M_k\}$ has a factorized form in steady state,
\be {\cal P}(\{M_k\})=\frac{\prod_{k=1}^{\nu} w(M_k)}{Z(M, V)}
\delta \left( {\sum_{k=1}^{\nu} M_k-M} \right) \label{PC1} \ee
where weight factor $w(M_k)$ depends only on mass $M_k$ of $k$th
subsystem and $Z(M, V) = Z(M, v \nu) = \prod_{k=1}^{\nu} [\int
dM_k w(M_k)] \delta({\sum_{k=1}^{\nu} M_k-M})$ the partition sum.

Probability distribution $P_v(m)$ of mass  $M_k=m$ in the $k$th 
subsystem of size $v$ is obtained by summing over all other subsystems 
$k'\ne k$, i.e.,
\bea
 P_v(m) = \frac{{w(m)} }{Z(M,V)}  {\tiny  \prod_{k'\ne k}\left[\int dM_{k'} 
w(M_{k'})\right] \delta\left({\sum_{k} M_{k}-M}\right)}.\nonumber
\eea
After expanding  $Z(M-m, V-v)$ in leading order of $m$ and taking
thermodynamic limit $M, V \gg 1$ with mass density $\rho=M/V$
fixed, we get \be P_v(m) = w(m) \frac{Z(M-m, V-v)} {Z(M, V)} =
\frac{w(m) e^{\mu(\rho) m}}{{\cal Z}(\mu)}, \label{Pmi} \ee where
${\cal Z}(\mu)=\int_0^{\infty} w(m) \exp(\mu m) dm$ and chemical
potential \be \mu(\rho)= \frac{d f(\rho)}{d\rho} \label{mu1} \ee
with $Z(M,V)=\exp[-Vf(\rho)]$ \cite{Zia_etal_JPhysA2004,
Zielen_JSP2002, BertinPRE2007, PradhanPRE2011}.
Using two equalities for mean of the subsystem mass $\langle m
\rangle=v\rho =\partial \ln {\cal Z}/\partial \mu$ and its
variance $\sigma_v^2 (\langle m \rangle)=(\langle m^2 \rangle -
\langle m \rangle^2) =
\partial^2 \ln {\cal Z}/\partial \mu^2$, a
fluctuation-response relation is obtained \be \frac{d \langle m
\rangle} {d \mu} = \sigma_v^2 (\langle m \rangle) \label{FR1}. \ee
For a homogeneous system, the mean and the variance should be
independent of $i$. Moreover, when mass is conserved, the variance
is a function of mean mass $\langle m \rangle$ or equivalently
density $\rho$. The analogy between Eq. \ref{FR1} and the
fluctuation-dissipation theorem in equilibrium is now evident.
Now Eqs. \ref{mu1}  and \ref{FR1}   can be integrated to obtain 
$Z(M, V)=\exp(-Vf(\rho))$  and then its  Laplace transform 
$\tilde{Z}(s, V) = \int_0^{\infty} Z(M, V) e^{-s M} dM.$ 
Since $[\tilde{Z}(s, V)]^{1/\nu}= \tilde w(s),$ the Laplace transform  
of $w(m),$ one can calculate   $w(m)$ straightforwardly and use it in 
Eq. \ref{Pmi} to get $P_v(m)$.

We demonstrate this procedure  explicitly  in  a specific case
where the variance of mass in a subsystem of size $v$ is
proportional to the square of its mean, i.e.,  \be \sigma_v^2 (\langle m
\rangle) \equiv \psi( \langle m \rangle)= \frac{\langle m \rangle^2}
{v\eta}, \label{Sigma1} \ee with $\eta$ a constant depending on 
parameters of a particular model. By integrating Eq. \ref{FR1} w.r.t. 
$\langle m \rangle = v \rho$ and using Eq. \ref{mu1}  we get \be 
\mu(\rho) = - \frac{\eta}{\rho} - \alpha ~{\rm  ; } ~ f(\rho) = - \eta 
\ln \rho - \alpha \rho - \beta. \label{mu2} \ee 
The integration constants $\alpha$ and $\beta$ do
not appear in the final expression of mass distribution.
Finally, we get the partition sum $Z(M,V) =
\exp[-Vf(\rho)]=({M}/{V})^{\eta V} \exp{(\alpha M + \beta V)}.$ Its
Laplace transform
$\tilde{Z}(s, V) = { e^{\beta V} \Gamma(\eta V +1)}/{[V^{\eta V}
(s-\alpha)^{^{(\eta V+1)}}] }$  can be written as \bea
\tilde{Z}(s,V) \simeq \frac{ e^{\beta V} \sqrt{2\pi \eta V} (\eta
V)^{\eta V} e^{-\eta V} } {V^{\eta V} (s-\alpha)^{^{(\eta V+1)}}}
= \frac{\rm{const.}}{ (s-\alpha)^{^{(\eta V+1)}} } \label{Zs} \eea
using asymptotic form of the gamma function $\Gamma(z+1)
\simeq \sqrt{2\pi z} z^z e^{-z}$
for large  $z$. The constant term in the numerator is independent of
$s$ and  thus $[\tilde{Z}(s, v\nu)]^{1/\nu}$ gives 
\bea 
\tilde w(s) = \frac{\rm{const.}}{ (s-\alpha)^{^{v \eta}} }
\label{Zsn} 
\eea 
in the thermodynamic limit  $\nu \rightarrow \infty.$ Consequently its 
inverse  Laplace  transform is  $w(m) \propto m^{v \eta-1} e^{\alpha m}.$
The  weight factor $w(m),$ along with Eqs. \ref{Pmi} and \ref{mu2}, 
leads to   $P_v(m)$ which is a  gamma distribution as in Eq. \ref{Pm1}.
This completes the proof  for  the functional form 
$\psi(x) \propto x^2.$  
The proof follows straightforwardly for discrete-mass models. Note 
that different classes  of mass  distributions $P_v(m)$ can  be generated 
for other functional forms of $\psi (x)$ (see Supplemental Material, 
section I). In all these cases, $P_v(m)$ serves as the large deviation 
function for mass in a large subsystem.

Though the  above proof  relies on  the strict factorization  condition Eq. 
\ref{PC1}, the  results 
are not that restrictive and are applicable to systems when the joint 
subsystem  mass distribution is {\it nearly} factorized. In fact, the 
near-factorization of the joint mass distribution  can be realized  in  
a wide class of systems 
as long as correlation length $\xi$  is  finite, i.e.,  spatial correlations 
are {\it not} long-ranged. In that case, subsystems  of size   much larger 
than $\xi$ can be  considered statistically independent and  
thus well described by  Eq. \ref{PC1} \cite{Lebowitz_JStatPhys1996, 
BertinPRL2006, PradhanPRL2010}.

{\it Models and Discussions.} --  We now illustrate the results in  the  
context of a broad class of mass-transport models where exact or near 
factorization  condition holds.
First we consider driven lattice gases (DLG) on a one dimensional ($1D$)  
periodic lattice of $L$ sites with discrete masses or
number of particles $m_i \in (0,1,2, \dots )$ at site $i$  where
the total mass $M$ is conserved. A particle hops  only  to its
right  nearest neighbor with rate $u(m_{i-1},m_i, m_{i+1})$ which
depends on the masses at departure site  $i$ and its nearest 
neighbors.  For a  specific rate $u(m_{i-1}, m_i, m_{i+1})=
{g(m_{i-1},m_i-1) g(m_i-1,m_{i+1})}/[g(m_{i-1},m_i)
g(m_i,m_{i+1})],$ the  steady-state  mass distribution of the 
model   is pair-factorized \cite{Evans_PRL2006}, 
i.e., ${\cal P}(\{m_i\}) \sim [\prod_{i=1}^{L} g(m_i,m_{i+1})]
\delta(\sum_i m_i -M)$.  
Unlike a site-wise  factorized   state,  i.e., Eq. \ref{PC1} with  $\nu=V$,  
the  pair-factorized steady state  does generate finite spatial correlations. 
For a  homogeneous  function  $g(x,y)= \Lambda^{-\delta} 
g(\Lambda x,\Lambda y),$  the   two-point  correlation    
for the rescaled mass $m'_i = m_i/\rho$ can be written as 
 $\langle m'_i m'_{i+r} \rangle \simeq  A(r)$   where 
\bea A(r) = \frac{ \prod_k \left[\int_0^{\infty} 
dm'_k g (m'_k, m'_{k+1}) \right] m'_i 
m'_{i+r} \delta \left(\sum_k m'_k - L \right)}{ \prod_k 
\left[\int_0^{\infty} dm'_k g(m'_k, m'_{k+1}) \right] \delta 
\left(\sum_k m'_k - L \right)} \nonumber
\eea 
is independent of $\rho$. The variance of mass $m=\sum_{i\in v} m_i$ in a  
subsystem  of size $v \gg 1$ can be calculated, ignoring  small 
boundary-corrections,  as $\sigma^2_v \simeq v \sum_{r=-\infty}^{\infty} 
(\langle m_i m_{i+r} \rangle - \rho^2) = {\langle m\rangle^2}/{\eta v}$
where $\eta^{-1}= \sum_{r=-\infty}^{\infty} [A(r)-1].$ Thus, in DLG   
with homogeneous   $g(x,y)$,   
$\psi(\langle m \rangle)$ is proportinal to $\langle m \rangle^2;$   
in fact this proportionality is generic in models  where  steady state is  
clusterwise factorized with $g$  a homogeneous function of  masses  at 
several sites (see  Supplemental Material, section II.B). In all these 
cases,  $P_v(m)$  should  be a gamma distribution.

\begin{figure}
\begin{center}
\leavevmode
\includegraphics[width=8.5cm,angle=0]{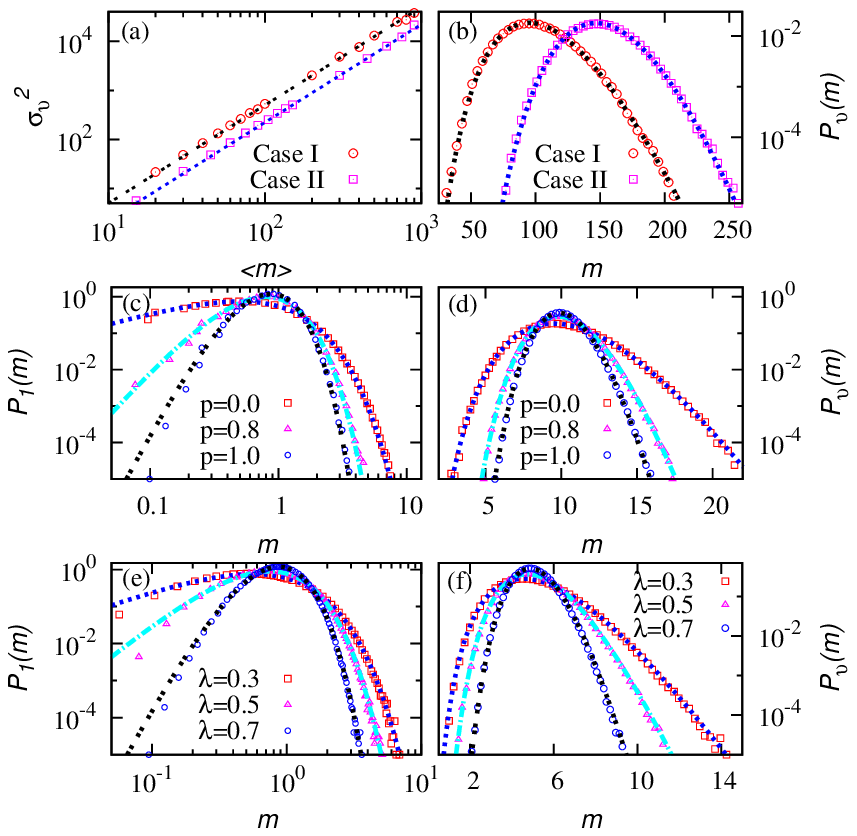}
\caption{ (Color online) {\it Driven lattice gases}: (a) Variance 
$\sigma^2_v$ of subsystem mass {\it vs.} its mean $\langle m \rangle$ 
(lines - fit to the form in Eq. \ref{Sigma1}) and (b) corresponding mass
distribution $P_v(m)$ for Case I. $\delta=1$,$c=0$ and $v=10$ (red 
circles) and Case II. $\delta=2$, $c=1$, $\alpha=1.5$ and $v=15$ (magenta 
squares). In both cases $\rho=10$ and $L=2000.$ {\it Mass chipping models}:  
Mass distribution $P_v(m)$ {\it vs.} mass $m$ with (c) $v=1$ and (d) 
$v=10$ for the 
model with $\lambda=1/2$ and $p=0$ (red squares), $0.8$ (magenta triangles) 
and $1$ (blue circles). {\it Wealth distribution models}: Mass distribution 
$P_v(m)$ {\it vs.} mass $m$ with (c) $v=1$ and  (d) $v=5$ for the model  with 
$\lambda=0.3$ (red squares), $0.5$ (magenta triangles) and $0.7$ 
(blue circles). In panels (c) - (f), $\rho=1$ and $L=1000$. Simulations - 
points, gamma distributions (Eq. \ref{Pm1}) - dotted lines.
}
\label{mass_transport}
\end{center}
\end{figure}

We now simulate DLG for two specific cases with $g(x,y)=(x^{\delta}+y^{\delta}
+cx^{\alpha}y^{\delta-\alpha})$ : Case I. $\delta=1$, $c=0$ and
Case II. $\delta=2$, $c=1$ and $\alpha=1.5.$ We then
calculate  the variance $\sigma^2_v \equiv \psi(\langle m\rangle)$ 
as a function of mean mass $\langle
m\rangle.$  As shown in Fig. \ref{mass_transport}(a),   in both the  cases, 
$\psi(\langle m\rangle) \propto \langle m \rangle^2$ as  in Eq. \ref{Sigma1}
with  $\eta \simeq 2.0$ and $\eta \simeq 3.0$ respectively.  For
these values of $\eta$, corresponding $P_v(m)$
obtained   from simulations are also  in excellent  agreement with
Eq. \ref{Pm1} as seen in Fig. \ref{mass_transport}(b).
Interestingly,  the value of  $\eta$ can be  calculated
analytically for case I where $\delta$ and  $\alpha$ are integers 
(see Supplemental Material, section II.A).

Next we consider a  generic variant of paradigmatic 
mass-transport processes, called {\it mass chipping models} (MCM)  
\cite{KrugGarcia2000, RajeshMajumdar2000, Zielen_JSP2002, Zielen_JSP2003,
Mohanty_JSTAT2012}.  These models are based on mass conserving dynamics
with linear  mixing  of  masses  at neighboring sites which 
ensures that $\sigma_v^2 \simeq \langle m\rangle^2/v\eta$ when  
the two-point correlations are negligible. Note that,  factorizability of  
steady state necessarily implies vanishing of two-point
correlations, but not {\it vice versa}.  However, when higher 
order correlations are also small, which is usually the case in these  
models, the steady state is nearly factorized and 
the resulting $P_v(m)$  can thus be well approximated by
gamma distribution for any $v$ (including $v=1$). We demonstrate 
these results considering  mainly the asymmetric mass transfer  in MCM; 
the symmetric case is then discussed  briefly.

In  $1D,$ asymmetric  MCM  is  defined  as follows. On a  
periodic lattice of $L$ sites with a mass variable $m_i\ge0$ at 
site $i,$ first $(1-\lambda)$  fraction of mass  $m_i$  is 
chipped off, leaving the rest of the mass  at $i$. 
Then a random fraction $r_i$ of the chipped-off mass 
$(1-\lambda)m_i$ is transferred to the right nearest neighbor 
and the rest comes back to site $i$.  At each site, the chipping 
process occurs with probability $p;$ thus the extreme 
limits $p=0$ and $1$ correspond respectively to random sequential 
(i.e., continuous-time dynamics) and
parallel updates. Effectively, at time $t$, mass $m_i(t)$ 
at site $i$ evolves following  a  {\it linear} mixing-dynamics
$m_i(t+1)=m_i(t) - (1-\lambda) [ \gamma_i m_i(t)
- \gamma_{i-1} m_{i-1}(t)],$ where $\gamma_i= \delta_i r_i$ with $\delta_i$ 
and $r_i$ are independent random variables drawn at
each site $i$ : $\delta_i=1$ or $0$ with probabilities $p$ and
$1-p$ respectively and $r_i$ is distributed according to a probability
distribution $\phi(r_i)$ in $[0,1]$. Ignoring two-point spatial
correlations, i.e., taking $\langle m_i m_{i-1} \rangle \approx
\langle m_i \rangle \langle m_{i-1} \rangle = \rho^2$, a very good
approximation in this case, the variance of mass  $\sigma_1^2
= \langle m_i^2 \rangle - \rho^2$ at a single site ($v=1$)  can be
calculated using  the stationarity condition
$\langle m_i^2(t+1) \rangle = \langle m_i^2(t) \rangle$. Then the variance
takes a simple form   $\sigma_1^2=\rho^2/\eta$ with
\be
\eta = \eta(\lambda,p,\mu_1,\mu_2) = \frac{\mu_1-(1-\lambda)\mu_2}
{(1-\lambda)(\mu_2-p\mu_1^2)}
\label{MCM_Var}
\ee
where  $\mu_k=\int_0^1 r^k \phi(r) dr$  moments of  $\phi(r).$ 
Moreover,  in these models, as   the two-point correlation function 
$\langle m_i m _{i+r}\rangle -\rho^2 \simeq 0$ vanishes
for  $|r|>0$, the variance of subsystem mass is given by $\sigma_v^2 \simeq
v \sigma_1^2 = \langle m \rangle^2/v\eta.$

A special  case  of  asymmetric MCM with
$\lambda=0$ and $p=1$ is  the `$q$' model of force
fluctuations \cite{Majumdar_Science1995, Majumdar_PRE1996}
which has a factorized steady state 
for a class of distribution $\phi(r)$ \cite{Zielen_JSP2002}.
In this case, $P_1(m)$ can be immediately obtained by using
$\eta={(\mu_1-\mu_2)}/{(\mu_2-\mu_1^2)}$ (from Eq. \ref{MCM_Var})
and $v=1$ in Eq. \ref{Pm1}. The mass distribution is in perfect
agreement with that obtained earlier \cite{Zielen_JSP2002} 
using generating function method. As a specific example, we
consider $\phi(r) = r^{a-1}(1-r)^{b-1}/B(a,b)$ with
$B(a,b)=\Gamma(a) \Gamma(b)/ \Gamma(a+b)$ for which the first two
moments  are  $\mu_1=a/(a+b)$ and
$\mu_2=ab/(a+b)^2(a+b+1)-a^2/(a+b)^2,$ and thus $\eta = a+b.$
Corresponding mass distributions is in agreement with that obtained in
\cite{Zielen_JSP2003}. For $\lambda=0$ and  $p<1$, the generalized 
asymmetric MCM becomes the asymmetric random average process
\cite{KrugGarcia2000, Zielen_JSP2002, Zielen_JSP2003}. We consider a 
specific case, when $r$ is uniformly distributed  in $[0,1]$,  the steady 
state is not  factorized and exact expression of  $P_1(m)$ 
is not known \cite{RajeshMajumdar2000}. However,  since the two-point 
correlations vanish \cite{RajeshMajumdar2000}, we  assume  the 
steady state  to be  nearly factorized  and  obtain  $P_1(m),$
a gamma distribution with  $\eta=2/(4-3p).$ We verified 
numerically  that this  simple  form agrees with  the actual $P_1(m)$ 
remarkably  well, except for small $m\ll \rho.$

For generic $\lambda$  and $p$ and for  a uniform $\phi(r)=1$ with 
$r \in [0,1]$,
the steady state is not factorized \cite{Mohanty_JSTAT2012} and
the spatial  correlations in general are nonzero. Consequently, no
closed form expression of the mass distribution is known,
except in a  mean-field approximation for $\lambda=1/2$ and $p=0$
\cite{Mohanty_JSTAT2012}. However, the spatial correlations are
small and gamma distribution provides in general a
good approximation of $P_v(m)$. In Fig. \ref{mass_transport}(c),
$P_1(m)$ versus $m$ is plotted for $\lambda=1/2$, $\rho=1$ and for
various $p=0, 0.8$ and $1$. One can see that $P_1(m)$ agrees quite
well with Eq. \ref{Pm1} with respective values of $\eta=2$, $5$,
and $8$. The deviation for $m \ll \rho$ is an
indication of the absence of strict factorization on the
single-site level. In Fig. \ref{mass_transport}(d), distribution
$P_v(m)$ of mass $m$ in a subsystem of volume $v=10$ is plotted as
a function of $m$ and it is in excellent agreement with Eq.
\ref{Pm1} almost over five orders of magnitude. Note that,
although Eq. \ref{PC1} does not strictly hold on the single-site
level, it holds extremely well for subsystems - a feature observed
in MCM or wealth distribution models (discussed later) for generic 
values of parameters.

In symmetric MCM's, with parallel update rules, a fraction
$\lambda$ of mass $m_i$ at site $i$ is retained at the site and
fraction $(1-\lambda)$ of the mass is randomly and symmetrically
distributed to the two nearest neighbor sites
\cite{Mohanty_JSTAT2012}: $m_i(t+1) = \lambda m_i(t) + (1-\lambda)
r_{i-1} m_{i-1}(t) + (1-\lambda) (1-r_{i+1}) m_{i+1}(t)$ where
$r_i$ uniformly distributed in $[0,1]$. For $\lambda=0$, the
steady state is factorized \cite{Mohanty_JSTAT2012} and
$P_1(m)$ is exactly given by Eq.
\ref{Pm1} with $\eta=2$. Clearly, when $\lambda=0$, both symmetric
and asymmetric MCM's with parallel updates result in $\eta=2$,
which explains why $P_1(m)$ in these two cases are the same
\cite{Mohanty_JSTAT2012}. Due to
the presence of finite spatial correlations, $P_1(m)$
with other update rules are not described by Eq. \ref{Pm1}.

Our results   are also applicable    
to models of energy transport \cite{KMP1982} and  wealth
distributions \cite{Patriarca_EPJB2010, Yakovenko,
CCModel, AnirbanC, Mohanty_PRE2006} defined on a $1D$ periodic
lattice of size $L$. Here,   $(1-\lambda)$ fraction of 
the sum  $m^{s}(t)= m_i(t)+m_{i+1} (t)$ of individual 
masses (equivalent  to  `energy' or `wealth') at nearest-neighbor 
sites $i$ and $i+1$   is  redistributed :
$ m_i(t+dt) =
\lambda m_i(t)  + r (1- \lambda) m^{s}(t)$ and $ m_{i+1}(t+dt) =
\lambda m_{i+1}(t) + (1-r) (1- \lambda) m^{s}(t)$ where $r$ is
uniformly distributed in $[0,1]$.  In this process  the total mass 
remains conserved. 
Assuming $\langle m_i m_{i-1} \rangle \approx 
\rho^2$, the variance is written
as $\sigma_1^2(\rho) \approx {\rho^2}/ {\eta(\lambda)}$ with
$\eta(\lambda)= {(1+2 \lambda)}/{(1-\lambda)}$, in agreement with
that found earlier numerically \cite{AnirbanC}. For $\lambda=0$, 
i.e., Kipnis-Marchioro-Presutti model in
equilibrium \cite{KMP1982}, the steady state is factorized and
$P_1(m)=\exp(-m/\rho)/\rho$  (with $\eta=v=1$)
is exact. For non-zero $\lambda$, as the spatial correlations are
small, the mass distributions, to a good approximation, are gamma 
distributions. In Fig. \ref{mass_transport}(e),
$P_1(m)$ versus $m$ is plotted for $\lambda=0.3$, $0.5$ and $0.7$
with $\rho=1$ and $L=1000$. Except for  $m \ll \rho$,  $P_1(m)$
agrees well with Eq. \ref{Pm1}. For a subsystem of size
$v=5$, the distributions $P_v(m)$, plotted in Fig.
\ref{mass_transport}(f) for the same parameter values as in the
single-site case, are in excellent agreement with Eq. \ref{Pm1}
for almost over five orders of magnitude.

{\it Summary.} -- In this Letter, we argue that  subsystem mass fluctuation
in driven systems, with mass conserving  dynamics and  short-ranged spatial
correlations, can be characterized from the functional dependence of
variance of subsystem mass on its mean. As  described in Eq. \ref{PC1}, 
such systems could effectively be considered as  a collection of  
statistically  independent subsystems of sizes much larger than correlation 
length, ensuring    existence of   an  equilibrium-like
chemical potential    and consequently   a fluctuation-response relation.
This   relation along with the functional form of the variance,
which can be  calculated  from the knowledge of only two-point
spatial correlations,  uniquely determines the  subsystem mass distribution.
We demonstrate  the result in  a broad class of mass-transport 
models where the variance of the subsystem mass is shown to be proportional 
to  the  square of its mean - consequently the mass distributions  are  
gamma distributions which have been observed in the past in different 
contexts. From a general 
perspective, this work could provide valuable insights in formulating a 
nonequilibrium thermodynamics for driven systems.

{\it Acknowledgment.} -- SC acknowledges the financial support from the
Council of Scientific and Industrial Research, India
(09/575(0099)/2012-EMR-I).

\end{document}